\documentclass[11pt]{article}
\usepackage{amsfonts,amssymb, graphicx,a4}

\newtheorem{theo}{Theorem}[section]

\newtheorem{prop}[theo]{Proposition}
\newtheorem{defi}[theo]{Definition}
\newtheorem{coro}[theo]{Corollary}

\begin{document}


\let\a=\alpha \let\b=\beta \let\c=\chi \let\d=\delta \let\e=\varepsilon
\let\f=\varphi \let\g=\gamma \let\h=\eta    \let\k=\kappa \let\l=\lambda
\let\m=\mu \let\n=\nu \let\om=\omega    \let\p=\pi \let\ph=\varphi
\let\r=\rho \let\s=\sigma \let\t=\tau \let\th=\vartheta
\let\y=\upsilon \let\x=\xi \let\z=\zeta
\let\D=\Delta \let\F=\Phi \let\G=\Gamma \let\L=\Lambda \let\Th=\Theta
\let\O=\Omega \let\P=\Pi \let\Ps=\Psi \let\Si=\Sigma \let\X=\Xi
\let\Y=\Upsilon

\global\newcount\numsec\global\newcount\numfor
\gdef\profonditastruttura{\dp\strutbox}
\def\senondefinito#1{\expandafter\ifx\csname#1\endcsname\relax}
\def\SIA #1,#2,#3 {\senondefinito{#1#2}
\expandafter\xdef\csname #1#2\endcsname{#3} \else
\write16{???? il simbolo #2 e' gia' stato definito !!!!} \fi}
\def\etichetta(#1){(\veroparagrafo.\veraformula)
\SIA e,#1,(\veroparagrafo.\veraformula)
 \global\advance\numfor by 1
 \write16{ EQ \equ(#1) ha simbolo #1 }}
\def\etichettaa(#1){(A\veroparagrafo.\veraformula)
 \SIA e,#1,(A\veroparagrafo.\veraformula)
 \global\advance\numfor by 1\write16{ EQ \equ(#1) ha simbolo #1 }}
\def\BOZZA{\def\alato(##1){
 {\vtop to \profonditastruttura{\baselineskip
 \profonditastruttura\vss
 \rlap{\kern-\hsize\kern-1.2truecm{$\scriptstyle##1$}}}}}}
\def\alato(#1){}
\def\veroparagrafo{\number\numsec}\def\veraformula{\number\numfor}
\def\Eq(#1){\eqno{\etichetta(#1)\alato(#1)}}
\def\eq(#1){\etichetta(#1)\alato(#1)}
\def\Eqa(#1){\eqno{\etichettaa(#1)\alato(#1)}}
\def\eqa(#1){\etichettaa(#1)\alato(#1)}
\def\equ(#1){\senondefinito{e#1}$\clubsuit$#1\else\csname e#1\endcsname\fi}
\let\EQ=\Eq
\def\GI{\mathbb{G}}
\def\VU{\mathbb{V}}
\def\vv{\vskip.2cm}
\def\vvv{\vskip.3cm}
\def\v{\vskip.1cm}
\def\FF{\mathcal{F}}

\def\sqr#1#2{{\vcenter{\vbox{\hrule height.#2pt
\hbox{\vrule width.#2pt height#1pt \kern#1pt
\vrule width.#2pt}\hrule height.#2pt}}}}
\def\square{\mathchoice\sqr34\sqr34\sqr{2.1}3\sqr{1.5}3}

\def\\{\noindent}
\def\EE{\mathbb{E}}
\def\Z{\mathbb{Z}}
\def\GG{\mathcal{G}}
\def\QQ{\mathcal{Q}}
\def\TT{\mathcal{T}}
\def\AA{\mathcal{A}}
\def\BB{\mathcal{B}}
\def\PP{\mathcal{P}}
\def\RR{\mathcal{R}}
\def\SS{\mathcal{S}}
\def\ES{\mathbf{S}}
\def\EP{\mathbf{P}}
\def\LL{\mathcal{L}}
\def\0{\emptyset}
\def\N{\mathbb{N}}
\def\setn{{\rm I}_n}
\def\CC{\mathcal{A}}
\def\E{\mathcal{E}}
\def\ER{{\bf R}}
\def\Lad{{\mathbb{L}^d}}
\def\Ed{{\mathbb{E}^d}}
\def\Zd{\mathbb{Z}^d}
\def\supp{{\rm supp}\,}

\def\arm{{}}
\font\bigfnt=cmbx10 scaled\magstep1


\title{Percolation on infinite graphs and isoperimetric inequalities}
\author{
Rog\'erio G. Alves$^{1,2}$\\
Aldo Procacci$^1$\\
Remy Sanchis$^1$\\
\\
\small{$^1$ Departamento de Matem\'atica UFMG}
\small{ 30161-970 - Belo Horizonte - MG
Brazil}
\\
\small{$^2$ Departamento de Matem\'atica UFOP}
\small{ 35400-000 - Ouro Preto - MG
Brazil}
}
\maketitle

\begin{abstract}
We consider the Bernoulli bond percolation process (with parameter $p$) on infinite graphs and
we give a general criterion for  bounded degree graphs to exhibit a non-trivial percolation threshold based either  on a single
isoperimetric inequality if the graph has a bi-infinite geodesic, or two isoperimetric inequalities if  the graph has not a bi-infinite geodesic. This new criterion extends previous criteria and brings together  a large class of
 amenable graphs (such as regular lattices) and non-amenable graphs (such trees).
We also study the finite connectivity in graphs satisfying the new  general criterion
and  show that graphs in this class with a bi-infinite geodesic always have finite connectivity functions with exponential decay as $p$ is sufficiently close to one.  On the other hand,  we show that there are
graphs in the same class  with no  bi-infinite geodesic for  which the finite connectivity decays  sub-exponentially (down to polynomially) in the highly supercritical phase even for $p$ arbitrarily close to one.

\end{abstract}

\numsec=1\numfor=1

\vskip.5cm
\section{Introduction}

Percolation is a subject which has been intensively  studied during the last decades mainly on the  $d$-dimensional unit cubic lattice $\Z^d$.
The study of   percolation processes
on infinite graphs other than $\mathbb{Z}^d$, started basically in the early nineties,
and has been focused essentially  on  non-amenable graphs (see e.g. \cite{H} and reference therein) and transitive graphs.
Some general
results and conjectures  about percolation on
infinite graphs
has been formulated  in the seminal paper
\cite{BS}  where the authors prove, among other results, that non-amenable graphs  do have a non-trivial percolation  threshold  (i.e. critical percolation probability
$p_c$ strictly less than 1).
In  \cite{BB} Babson and Benjamini have introduced a parameter   depending on the graph's structure  which should be relevant in percolation.
Basically, given an infinite graph $G=(V,E)$,  this parameter is the minimum  $t\in\mathbb{N}$ such that any minimal cut set in $G$  is  $t$-close, i.e. it is connected in the  graph $G^t$ which has  the same vertex set $V$ of $G$
and  edge set  formed by those pairs $\{x,y\}\subset V$ whose  distance is less or equal than  $t$. A graph for which $t$ is finite is said to have the
quasi-connected minimal cut sets property (see e.g. \cite{Sc}). Babson and Benjamni showed, via Peierls argument, that if $t<\infty$ then
$p_c<1$ for  Cayley graphs of finitely presented infinite groups which are not
a finite extension of $\mathbb Z$ (most of the regular lattices fall in this class).
Later,  Procacci and Scoppola  \cite {PS7} pointed out that $t<\infty$ was a sufficient condition for $p_c<1$ in a large class of bounded degree graphs,namely,
graphs with a bi-infinite geodesic or satisfying a very mild isoperimetric inequality. So,
despite that most of  the non-amenable graphs has $t=\infty$ (but $p_c<1$!), there was hope that the finiteness of $t$ could still be a key information to establish whether
$p_c<1$,  at least for a bounded degree
{\it amenable} graph. In particular,  Babson and Benjamini  explicitly conjectured that the finiteness of $t$
could be a sufficient condition for $p_c<1$   for all amenable quasi-transitive graphs with one end.
However recently   Tim\'ar \cite{T} provided
a counterexample, i.e.     a one-ended transitive amenable graph with $p_c<1$ (the  Diestel-Leader graph $DL(2,2)$)
for which the Babson-Benjamini parameter $t$  is infinite.
 This, together with the fact that most of the  non-amenable graphs (e.g. trees)
have $t=\infty$  and  $p_c<1$,   seems to suggets that  $t$  may not be the right quantity to look at in order to implement Peierls argument in general graphs.

In conclusion, even
speculating that a necessary and sufficient criterion  for a general graph to have $p_c<1$ is probably too much to ask, it would
be desirable to obtain at least a sufficient criterion able to include  as much as possible  graphs with known  non-trivial percolation.

The results presented in this note  are a  little step in  this direction. We present
in fact a new sufficient criterion for a graph to have $p_c<1$ which now brings together   both amenable and  non-amenable graphs (including
graphs with Babson-Benjamini parameter $t=\infty$).
The new criterion is based   on a single
isoperimetric inequality if the graph has a bi-infinite geodesic, while two isoperimetric inequalities are required if
the graph has not a bi-infinite geodesic.

We also study the two-point connectivity function in graphs satisfying the general criterion above.
In particular we investigate the possibility, originally discussed  in \cite{PS8}, about the existence of percolation processes in infinite graphs  for which the finite connectivity
decays  non-exponentially
even in the highly supercritical phase. Indeed, given a graph $G$ without bi-infinite geodesics which falls in the class satisfying our new criterion, we show that the decay of  the finite connectivity may be not exponential, presenting an example.
On the other hand, if $G$ has a bi-infinite geodesic, then we show that the connectivity functions decay exponentially in the highly supercritical phase.

The paper is organized as follows. In  Section 2 we rapidly review some definitions about graphs and  remind the basic notions of the Bernoulli bond percolation process in graphs, defining in particular the finite connectivity functions
and the critical percolation probability. In Section 3 we present our results in form of four theorems. Finally,
in Section 4 we give the proofs of these theorems.

\numsec=2\numfor=1
\section{Notation and Definitions: graphs and percolation}

\subsection{Infinite graphs}\label{graphs}
Throughout the paper, whenever  $X$ is a set, we will denote by $|X|$ its cardinality. Let $G=(V,E)$ be a graph with vertex set $V$ and edge set $E$. If $x,y\in V$, we denote by $d_G(x,y)$ the usual path distance in $G$.
If $W\subset V$, let $G[W]=(W,E[W])$ denote the induced subgraph where $E[W]=\{\{x,y\}\in E: x\in W, y\in W\}$.
A set $W\subset V$ is {\it connected} in $G$
if $G[W]$ is connected.


Given $G=(V, E)$ connected and $W\subset V$, we denote {\it the edge boundary} of $W$ by
$\partial_e W= \{e\in E: |e\cap W|=1 \}$.
We denote
$\partial^{\rm ext}_v W= \{x\in V\backslash W: d_G(x,W)=1 \}$ {\it the vertex external boundary} of $W$
and
$\partial^{\rm int}_v W= \{x\in W: d_G(x,V\backslash W)=1 \}$ {\it the vertex internal boundary} of $W$.
We denote $W^{c}= W\cup\partial^{\rm ext}_v W$ the {\it closure of $W$}. Note that $W^c$ is connected and $E[{W^c}]=E[W]\cup\partial_e W$.
We finally denote
${\rm diam}(W)=\max_{x,y\in W}d_G(x,y)$.

Throughout the paper, the symbol $\GI=(\VU,\EE)$ will always denote a graph which is {\it infinite, connected and bounded degree}, with
$\D(\GI)\equiv \max_{v\in \VU}|\partial_ev|<\infty$ being its  maximum degree.
We denote by $\CC_\GI=\{W\subset \VU: |W|<\infty~{\rm and}~\GI[W]{\rm ~is~ connected}\}$ the set
of all finite and connected subsets of vertices of  an infinite graph $\GI$.


A {\it geodesic ray} $\r=(V_\r, E_\r)$ in $\GI$ is an {\it infinite} sub-graph of $\GI$ such that
$
V_\r=\{x_0, \dots , x_n, \dots\}$,
$E_\r=\left\{\{x_0, x_1\}, \dots , \{x_{n-1},x_n\},\dots\right\}$ and $d_\GI(x_0, x_n)=n$ for all $n\in \mathbb{N}$.
Let $\r$ and $\r'$ be two geodesic rays in $\GI$, both starting at $x_0$,
with vertex sets $V_\r=\{x_0,x_1, \dots , x_n,\dots\}$
and $V_{\r'}=\{x_0, y_1, \dots , y_n,\dots\}$ respectively. If
$V_\r$ and $V_{\r'}$ are such that
$d_\GI(x_n,y_m)=n+m$ for any $\{n,m\}\subset \mathbb{N}$,
then the union $\r\cup \r'=(V_\r\cup V_{\r'}, E_\r\cup E_{\r'})$ is called
a {\it bi-infinite geodesic} in $\GI$.

A finite connected graph $\t=(V,E)$ is a {\it tree} if $|E|=|V|-1$. If $G$ is a graph, we denote by $\TT(G)$ the set of all subgraphs of $G$ which are trees.




\begin{defi}
Given  a graph $\GI=(\VU,\EE)$,  and given a connected set  $W\in \CC_\GI$,
we define
the tree distance $d_\GI^{\rm t}(\partial_eW)$ of the edge-boundary $\partial_eW$ of $W$ as
$$
d_\GI^{\rm t}(\partial_eW) = \min_{(V,E)\in \TT({\GI[W^c]})\atop \partial_e\subset E}|E| \Eq(dgte0)
$$
\end{defi}
\\{\bf Remark}. Given $W\in\CC_\GI$, let $\ell_{\partial_eW}$  be the  number of egdes of  $\EE[W]$ necessary to connect  $\partial^{int}_v W$, then
$d_\GI^{\rm t}(\partial_eW)$ can also be written as
$$
d_\GI^{\rm t}(\partial_eW)=  \ell_{\partial_eW} + |\partial_eW| \Eq(dgtew)
$$
Hence, since $\ell_{\partial_eW}$ is at most $|W|-1$, we get immediately the following upper and lower bounds for $d_\GI^{\rm t}(\partial_eW)$
$$
|\partial_eW|\le d_\GI^{\rm t}(\partial_eW)\le  |\partial_eW| +|W|-1 \Eq(dgtewle)
$$
where the equalitiy  $|\partial_eW|= d_\GI^{\rm t}(\partial_eW)$ holds if and only if  $\partial_eW$ is connected and the equality
$|\partial_eW|= |\partial_eW| +|W|-1$ holds if and only if $\GI[W]$ is a tree.
\vv
A {\it cut set} of a graph $G=(V,E)$ is
a set $\g\subset {E}$ such that the graph $G\backslash\g\equiv(V, E\backslash \g)$ is disconnected.

\def\OO{{\cal O}}
\begin{defi}\label{due1}
Given  $\GI=(\VU,\EE)$,
a finite cut set $\g\subset \EE$
is called a contour if
$\GI\backslash\g$ has exactly one finite
connected component and is minimal with respect to this property, i.e.  for all edges $e\in \g$ the graph $(\VU, \EE\backslash(\g\backslash e))$ has no finite connected component.
We denote by $\FF_{\GI}$ the set of all contours in $\GI$.
\end{defi}
This definition
generalizes in some sense the  notion of Peierls contour in the Ising model. These objects were called ``$(v,\infty)$- minimal cut set" in \cite{BB}, ``Peierls contours" in \cite{PS7}, and ``fences" in \cite{PS8}.
It is worth to mention some recent extensions
of  the general notion of  Pirogov-Sinai contours  for trees in \cite{Ro} and \cite{GRS}.

If $\g$ is a contour in $\GI$, we denote by $G_\g=(I_\g , E_\g)$ the
unique finite connected component of $\GI\backslash\g$; the set $I_\g\subset \VU$ is called {\it the vertex interior of the contour $\g$} and the set $E_\g\subset \EE$ is called {\it the edge interior of the contour $\g$}. 
We denote by $G_\g^c=(I^c_\g , E^c_\g)$ the graph with vertex set $I^c_\g=I_\g\cup\partial^{\rm ext}_v I_\g$ and edge set $E^c_\g=E_\g\cup\g$ and call it {\it the closure of $G_\g$}.




\vv
\\{\bf Remark}.
If $\g\in \FF_\GI$ is a contour, then $I_\g\in\CC_\GI$ and $\partial_e  I_\g=\g$. So  the tree distance $d_\GI^{\rm t}(\g)$ of the contour $\g$ is
$$
d_\GI^{\rm t}(\g) = \min_{(V,E)\in \TT({G^c_\g})\atop \g\subset E}|E|= \ell_\g+|\g| \Eq(dgte)
$$
where  $\ell_\g$  is the  number of egdes of  $E_\g$ necessary to connect  $\partial^{int}_v I_\g$.

\vskip6.0cm

\vv

\\Given a contour $\g$ in $\GI=(\VU,\EE)$ and a set of vertices
$X\subset \VU$, we say that $\g$ surrounds $X$ and we write
$\g\odot X$ if $X\subset I_\g$. We say that $\g$ separates $X$ and we write
$\g\otimes X$ if $0<|X\cap I_\g|<|X|$.

We denote  by $\FF^n_\GI$ the set of all contours with cardinality $n$, by $\FF_\GI(X)$  ($\FF^n_\GI(X)$) the set of $\g\in \FF_\GI$ ($\g\in \FF^n_\GI$) such that $\g\odot X$ and finally, with a slight abuse of notation, for $e\in E$, we denote by $\FF^n_\GI(e)$ the set of contours $\g\in\FF^n_\GI$ such that $e\in \g$.

We now introduce two isoperimetric constants, $R_\GI$ and $P_\GI$,   in a bounded degree graph $\GI$, which play a central role
in all our results.
\begin{defi}
Given $\GI=(\VU,\EE)$, let
$$
R_\GI=\inf_{W\in  \CC_\GI} {|\partial_e W|\over d^{\rm t}_{\GI}(\partial_e W)} \Eq(iso1)
$$
and
$$
P_\GI=\inf_{W\in  \CC_\GI} {|\partial_e W|\over \log({\rm diam}(W))}\Eq(iso2)
$$
\v
\\We call $R_\GI$ the ``contour constant" of $\GI$ and $P_\GI$ the ``wedge constant" of $\GI$.
\end{defi}

The  wedge constant $P_\GI$ of a graph $\GI$ was explicitly introduced  in \cite{PS7}  and   its role, as far as percolation  on $\GI$ is concerned,
has already been pointed out there. See also \cite{Gr} and \cite{CC}  where  percolation in  subsets of $ \mathbb{Z}^d$ with wedge growing logarithmically with the diameter has originally been considered.

The contour constant $R_\GI$, as far as we know,   is rather new in the literature. A related quantity has been recently
introduced by Campari and Cassi \cite{CaC} in the study of the Ising model on general graphs.
In \cite{CaC} the authors define a finite constant $l$ depending on the structure of the graph $\GI$, such that
any contour in $\GI$ (according to definition \ref{due1}) with cardinality $n$ is
connectable with no more than $l ·n$ vertices. It is easy to see that $l$ is essentially the inverse of $R_\GI$. Indeed, if $R_\GI>0$, then for any contour $\g$ , $d^{\rm t}_{G}(\g)\le {1\over R_\GI}|\g|$, or, in other word,  recalling  \equ(dgte), $\g$ is connectable (by a tree wich contains $\g$) using at most ${1\over R_\GI}|\g|$ edges. So an easy computation yields
${1\over R_\GI}-1\le l\le {1\over R_\GI}+1$.

The definition of the contour constant $R_\GI$ as  formulated in \equ(iso1) resembles
that of the more usual and known  {\it Cheeger  constant} $C_\GI$, a.k.a. {\it isoperimetric constant}  or  {\it expansion constant}, defined (see e.g. \cite{BS} or \cite{CPP}) as
$$
C_\GI=\inf_{W\in \CC_\GI}{|\partial_e W|\over |W|} \Eq(chee)
$$
However,  we want to stress that the behavior of $R_\GI$ is  quite different from that of $C_\GI$ as $\GI$ varies in the class of infinite graph.
Indeed,  by  inequality \equ(dgtewle),  it is easy to see  that

$$
R_\GI\ge {C_\GI\over C_\GI+1}\Eq(rgcg)
$$
so that
$R_\GI$ is positive whenever the Cheeger constant of $\GI$ is positive.
On the other hand, the converse is not  true: the positivity of  $R_\GI$ does not imply, in general, that of the Cheeger constant. In particular,
 $R_\GI$ is strictly positive in all amenable graphs (i.e. graphs with $C_\GI=0$) for which the Babson-Benjamini parameter
$t$ is finite, since, by definition \equ(dgtew) and by definition of $t$ given in the introduction, we have that $d^{\rm t}_{\GI}(\partial_e W)\le (1+t)|\partial_e W|$ and hence
$$
R_\GI\ge {1\over t+1}\Eq(rgt)
$$
The unit cubic lattice $\mathbb{Z}^d$ (for $d>1$)   is a topical example of a graph
with  $C_\GI=0$,  $t$ finite, and  $R_\GI>0$.

Finally, using the concept of contours, we  introduce a new kind of distance between two vertices $x,y$ in a graph $\GI$.
\begin{defi}\label{non root}
Given a graph  $\GI=(\VU, \EE)$, let $x,y\in \VU$.
We define the {\it contour distance} $f_\GI(x,y)$ between $x$ and $y$ by
$$
f_\GI(x,y)=\min_{\g\in \FF_\GI\atop \g\bigodot \{x,y\}}|\g| \Eq(fGI)
$$
\end{defi}
As we will show ahead, the contour distance can play an important role in the decay properties of the truncated connectivity functions of a graph  in the supercritical phase.

\subsection{Independent Percolation on infinite graphs}
Given $\GI=(\VU,\EE)$ and $p\in [0,1]$, we associate to each
edge $e\in \EE$ i.i.d. Bernoulli variables $\om(e)$,
taking the value $\om(e)=1$ (meaning that the edge $e$ is open) with probability
$p$, or else the value $\om(e)=0$ (meaning that the edge $e$ is closed) with probability $1-p$. Let $P_p$ denote the standard product measure on the configurations of edges in $\GI$.
A configuration
$\om$ of the process is a function $\om: \EE\to \{0,1\}: e\mapsto \om(e)$. We call $\O_\GI$ the set of
configurations in $\GI$. Given $\om\in \O_\GI$ we denote by
$O(\om)$ the subset of $\EE$ given by $O(\om)=\{e\in \EE: \om(e)=1\}$
and by $C(\om)$ the
set $C(\om)=\{e\in \EE: \om(e)=0\}$.

If $G_N=(V_N,E_N)$ is a finite subgraph of $\GI$, let $\O_{N}$ be the of configurations in $G_N$, and let $\om\in\O_N$, then the
the probability
$P_p(\om)$ is given explicitly by

$$
P_p(\om)=
p^{|O(\om)|}(1-p)^{|C(\om)|}\Eq(restr)
$$

Given a configuration $\om\in \O_\GI$, an {\it open cluster} $g$ of $\om$ is
a connected subgraph $g=(V_g, E_g)$ of $\GI$ such that
$\om(e)=1$ for all $e\in E_g$, and $\om(e)=0$ for all $e\in \partial g$ where $\partial g=\{e\in \EE: |e\cap V_g|=1\}$ is the external edge boundary of $g$. If $g=(V_g, E_g)$ is an open cluster
of a given configuration $\om$
and $X$ is a non-empty subset of $V_g$ we write shortly $X\subset g$, and write shortly $|g|$ in place of $|E_g|$.

\begin{defi}\label{connec}
Let $\GI=(\VU,\EE)$ be an infinite graph and let $X\subset V$ such that $|X|=n$.
The $n$-point finite connectivity function $\phi^{\rm f}_{p}(X)$ is defined as
$$
\phi^{\rm f}_{p}(X)= P_{p}(\exists \;{\rm open \;cluster}\; g: X\subset g,\, |g|<\infty)\Eq(conneff)
$$
\end{defi}

In this paper we will be interested only in $two$-points finite correlations and {\it one}-point finite correlation (i.e $|X|\le 2$). In particular, the one-point finite correlation is directly related
to the percolation probability.
Given a graph $\GI=(\VU,\EE)$ and a vertex $x\in \VU$, the percolation
probability, i.e. the probability that
there is an infinite open cluster passing through $x$, is defined as
$$
\theta_p(x)=P_p(\exists \;{\rm open \;cluster}\; g: \{x\}\subset g,\, |g|=\infty)
$$
By \equ (conneff),  $\theta_p(x)$ may be written in term
of connectivity functions as
$$
\theta_p(x)=
1- \phi^{\rm f}_{p}(x).\Eq(theta)
$$

A standard coupling argument shows that, in any  graph $\GI$, $\theta_p(x)$ is an increasing function of $p$ (see e.g. \cite{G1}, Theorem 2.1). And,  as $\GI$ is connected, the critical percolation probability $p_c$ does not depend on the choice of $x$, since if
$\theta_p(x)>0$, then, by FKG, $\theta_p(y)>0$ for any two vertices $x,y$.

The critical percolation probability $p_c(\GI)$ for the graph
$G$ is defined by
$$
p_c= \sup_{p\in [0,1]}
\{p: \theta_p(x)=0\}\Eq(pc)
$$
and we say that the system is in the {\it subcritical phase} if $p\in [0,p_c)$ and in the {\it supercritical phase} if $p\in (p_c,1]$.



A way to show that $p_c<1$ is to establish a non-trivial upper bound for the one-point correlation
$\phi^{\rm f}_{p}$, and this can be obtained via the so called Peierls argument. This very famous tool was originally stated for the Ising model on $\Z^2$
in the low-temperature phase, but, once the notion of contours (according to  def. \ref{due1}) is introduced, the argument can be generalized for bond Bernoulli percolation in any graph $\GI$. With the notations and definitions previously introduced,
the Peierls argument for a general  graph $\GI=(\VU,\EE)$ can be stated as follows.
\begin{prop}[Peierls argument]\label{peia}
Let $\GI=(\VU,\EE)$ be an infinite graph. If it is possible to find a finite positive constant $r$ and a vertex $x\in \VU$ such that, for all $n\in \mathbb{N}$,
$$|\FF^n_\GI(x)|\le r^n,\Eq(pea)
$$
then
$$
p_c\le 1-{1\over 2r}
$$
\end{prop}
Such a proposition follows immediately by observing that if $g=(V_g,E_g)$ is finite open cluster such that $\{x\} \subset V_g$, then there is a contour $\g$ such that
$\g\odot x$ and $\g\subset \partial g$, i.e. $\g$ is formed by closed edges. Therefore,
$$
\phi^{\rm f}_{p}(x)
\le \sum_{n\ge 1}\sum_{\g\in \FF_\GI^n(x)}(1-p)^{|\g|}= \sum_{n\ge 1}[r(1-p)]^n <1\Eq(inn)
$$
as soon as $p> 1-{1\over 2r}$; whence $p_c\le 1-{1\over 2r}$.\hfill$\Box$

In the next section we present our results in form of four
 theorems. The first two, Theorem \ref{criton} and Theorem \ref{criton2}, concerns non triviality of percolation threshold $p_c$ in infinite graphs
 with finite maximum degree and provides new criteria to establish, via a sufficient condition,
if a graph $\GI$ in this class has $p_c<1$.  Theorems \ref{geoteo} and \ref{supf} will  concern the decay of  the connectivity function for  bounded degree graphs.

\numsec=3\numfor=1
\section{Results}

\subsection{New sufficient criteria for a graph $\GI$ to have $p_c<1$}

In order to implement the Peierls argument in a graph $\GI$ one must be able to bound exponentially the number of contours of cardinality $n$ that surrounds a fixed vertex.
This is relatively easy if $G$ is the cubic lattice $\Z^d$ but it may be a desperate task for general graphs.
In their seminal paper \cite{BS}, Benjamini and Schramm wondered whether is possible
to replace in a general graph the Peierls condition \equ(pea) by some more friendly isoperimetric inequality.

Our first result here below may be viewed, in our opinion, as a step in this direction.

 \begin{theo}\label{criton}
Let $\GI=(\VU,\EE)$ be an infinite graph with maximum degree $\D$ and with countour constant, defined in \equ(iso1), equal to $R_\GI$. Suppose that
\vskip.3cm
{\it i) $R_\GI>0$}
\vskip.2cm
{\it ii)}  $\GI$ has a bi-infinite geodesic
\vskip.3cm
\\Then $\GI$ has a non trivial percolation threshold and
$$
p_c(\GI)
\le
1- {1\over 2(2e\D^2)^{1/R_\GI}}\Eq(pci)
$$
\end{theo}
Observing that  any  quasi-transitive graph has  always a bi-infinite geodesic
(see e.g. proposition 5.2 in \cite{TW} or  lemma 5.7 of \cite{DW}), Theorem \ref{criton} immediately implies the following corollary.

\begin{coro}\label{coro32}
A quasi-transitive graph $\GI$ such that $R_\GI>0$  has $p_c<1$.
\end{coro}
\\{\bf Remark 1}.
Theorem \ref{criton} is a genuine extension of  Theorem 1 of \cite{PS7}.
In particular, by \equ(rgcg) and \equ(rgt), one immediately sees that the class of graphs satisfying the hypothesis of Theorem \ref{criton} contains
all graphs with positive Cheeger constant and all graphs with Babson-Benjamini paramer $t$ finite. So, for example, a regular tree, for which $t$ is infinite and $C_\GI>0$, as well as a regular amenable lattice,
for which  $C_\GI=0$, are both in the class of graphs satisfying Theorem \ref{criton}.
We would also draw the attention to the possibility that  $R_\GI$ may  play  a role
similar (or alternative)
to that of the isoperimetric dimension $d_{iso}$ (see e.g. \cite{BS} or \cite{CaC} for  its definition).
In particular, it seems interesting to inquire whether
$R_\GI>0$ for graphs with
$d_{iso}>1$, for in this case  Theorem \ref{criton} could be regarded as generalization, at least for graphs with a bi-infinite geodesic, of the conjecture state in \cite{BS}  (see there question 3.4).

\vv





\vv

\\{\bf Remark 2}. Concerning Corollary \ref{coro32}, we recall   that every  quasi-transitive infinite graph $\GI$  has either 1 end, or two ends, or infinitely many ends  (see e.g. \cite{HLS} at the end of sec. 8 and references therein).
If  $\GI$  has infinitely many ends, then $\GI$ is non-amenable (see again \cite{HLS} or \cite{Mo}, proposition 6.2), hence  $C_\GI>0$  and, by \equ(rgcg),
$R_\GI>0$. So $p_c<1$. If $\GI$ has two ends,
then $\GI$ is a finite extension of $ \mathbb{Z}$  and hence $p_c=1$. We are thus left with the quasi-transitive amenable graphs with one end, for which $R_\GI>0$
is a sufficient condition for $G$ to have $p_c<1$. As pointed in by Tim\'ar in \cite{T}, the Diestel-Leader graph $DL(2,2)$  is an  example of amenable (hence with  Cheeger constant equal to zero) transitive graph with one end  exhibiting  contours which are not $t$-closed (in the Babson-Benjamini sense) for any $t\in \mathbb{N}$, such contours being the edge boundaries of the so-called tetrahedrons $T_n$.
On the other hand, one can check immediately from see definition of  $T_n$ (see Definition 2 in  \cite{BW}) that $|\partial T_n|=2^{n+2}$ and  $d^{\rm t}_{DL(2,2)}(\partial_e T_n)= 2\sum_{k=1}^n 2^k$, so that $d^{\rm t}_{DL(2,2)}(\partial_e T_n)/|\partial T_n| \le 1$, i.e.
tetrahedrons  in  $DL(2,2)$ do satisfy \equ(iso1). So the question raised by Babson and Benjamini (\cite{BB} question 3) and answered negativally by Tim\'ar can be replaced by the following.

\vv
\\{\bf Question}. {\it
Are there a one-ended quasi-transitive amenable graph $\GI$ for which $R_\GI$ is zero?}
\vv
\\Our second  result  refers to the  class of graphs with no bi-infinite geodesic for which Theorem \ref{criton} cannot be applied.

\begin{theo}\label{criton2}
Let $\GI=(\VU,\EE)$ be an infinite graph with maximum degree $\D$ with countour constant, defined in \equ(iso1), equal to $R_\GI$ and  wedge constant, defined in \equ(iso2) equal to $P_\GI$. Suppose that
\vskip.3cm
{\it i) $R_\GI>0$}
\vskip.2cm
{\it ii)}  $P_\GI>0$

\vskip.3cm
\\Then $\GI$ has a non trivial percolation threshold and
$$
p_c(\GI)
<
1- {1\over 2(2\D^2)^{1/R_\GI}e^{1/P_\GI}}\Eq(pci2)
$$
\end{theo}

\\{\bf Remark 3}.  Theorem \ref{criton2} can be directly compared with the results recently  obtained by Campari and Cassi in \cite{CaC},
which can be resumed as follows (see there the theorem at pag. 021108-4).

\\{\it For all graphs with isoperimetric dimension
$d_{iso}>1$  and contours (of cardinality $n$) which are connectable with no
more than $l ·n$ vertices  (where $l$ is a finite constant), the number of contours with cardinality $n$ surrounding a fixed vertex $x$ is bounded by
${C}^n$ with $C$ constant.}

\\As previously discussed  the condition $l$ finite is equivalent to require  $R_\GI>0$, however the first Campari-Cassi condition, i.e., $d_{iso}>1$,  is much stronger than requiring   simply $P_\GI>0$. Indeed,
by  definition, the isoperimetric dimension of $\GI$ is greater than one
 if  and only if there exists an $\e>0$ such that $\inf_W |\partial_e W|/|W|^{\e}>0$. This clearly implies $P_\GI>0$, since, for all $\e>0$,
$\log({\rm diam}(W))\le |W|^{\e}$ as soon as $|W|$ is sufficiently large.  As an example,  consider the graph $\GI=(\VU,\EE)$ whose  vertex set $\VU$ is the subset  of $\Z^2$ given by $\VU=\{(n_1,n_2)\in \Z^2: n_1\ge0~{\rm and}~0\le n_2\le \ln(1+n_1)\}$ and  whose edge set is formed by the nearest neighbors in $\VU$. This graph has $d_{iso}=1$ so it is outside the class
 of graphs considered in \cite{CaC} but it has both $P_\GI>0$ and $R_\GI>0$.

\subsection{Two-point finite connectivity in bounded degree graphs}

In \cite{PS8} the authors obtained  (see there Theorem 4.1) an upper bound for   the two-point connectivity function of the Random Cluster Model with parameters $p\in [0,1]$ and $q>0$ on a bounded degree graph $\GI$, showing that this function decays at least exponentially in $\GI$ as soon as, for fixed $q$,  $p$ is sufficiently  close to zero (i.e. in the highly subcritical phase).
Since  the independent percolation process on a graph  $G$ coincides with Random Cluster Model on the same graph with parameter $q=1$, Theorem  4.1. of \cite{PS8}
immediately implies the following well known result.
\vv
\\{\bf Claim}. {\it In any graph $\GI$ with maximum degree $\D$, as far as independent percolation with parameter $p$ on $\GI$ is considered,
the two point  connectivity function  $\phi^{\rm f}_{p}(x,y)$ always decays (at least) exponentially, as $d_\GI(x,y)\to \infty$,  if  $p$ is sufficiently  small}.
\vv
A proof of this statement specifically for percolation  can be found e.g. in  \cite{Gr}  (first part of the proof of Theorem 1.10). The argument there is performed for $\mathbb{Z}^d$ but
 it can be easily generalized for any bounded degree graph. We also mention that this  claim has actually  been proved to be true   in  the whole subcritical regime
for  special classes of graphs, namely,   $\mathbb{Z}^d$ and regular lattices (see  \cite{AN}, section 5.3),  quasi-transitive graphs (see \cite{AV}, Theorem 3) and
non-amenable graphs (see \cite{Sc}, Theorem 5.3).

In the same paper \cite{PS8}, authors also studied the  decay properties of the two-point connectivity function
of the Random Cluster Model with parameters $p\in [0,1]$ and $q>0$ on $\GI$
when $p$ is close to 1 (i.e. in the highly supercritical phase). In this case, however, they obtained (Theorem 5.9 of \cite{PS8})
an upper  bound for  the connectivity functions of the form ($C_1, C_2$ are constants)
$$
C_1\exp\{-C_2\,f_\GI(x,y)\}\Eq(lbo)
$$
where  the function $f_\GI(x,y)$ is
the contour  distance  $f_\GI(x,y)$ defined in \equ(fGI).
Now, it easy to provide examples of graphs for which  $f_\GI(x,y)/d_\GI(x,y)\to 0$ as $d_\GI(x,y)\to\infty$. So, being  able get a lower bound of the same form of \equ(lbo),  one  could raise the
question (raised in fact in \cite{PS8}) whether  there are graphs for which the finite connectivity functions decay sub-exponentially, even  for $p$ arbitrarily close to 1.

In this last section, motivated by the bounds obtained in Theorem 5.9 of \cite{PS8}, we  present two  theorems  concerning the decay of  the connectivity function for   Bernoulli
percolation in a bounded degree graph $\GI$  in the supercritical phase.

In the first one, Theorem \ref{geoteo},
we show that the exponential decay of connectivities also holds in the supercritical phase, as  $p$ is sufficiently close to one,
{\it if one restrict himself to the class of   graphs satisfying  the hypothesis of Theorem \ref{criton}}, i.e.   graphs with a bi-infinite geodesic
and with positive contour constant $R_\GI>0$.

Our second result, Theorem \ref{supf}, concerns the decay of connectivity functions in bounded degree graphs { \it satisfying this time the hypothesis of Theorem \ref{criton2}},
and
essentially confirms the possibility raised in \cite{PS8}. Namely, a sub-exponential decay of the connectivity functions  may indeed occur in a  graph $\GI$ with positive contour constant even for $p$ arbitrarily close to one
(see the example below),
but only if  $G$ has no bi-infinite geodesic.

\begin{theo}\label{geoteo} Let $\GI=(\VU,\EE)$ a bounded degree graph which satisfies the hypothesis of Theorem \ref{criton}, i.e.  $\GI$ has a bi-infinite geodesic and
$R_\GI>0$. Then, as soon as $p\ge {4r\over 4r+1}$ and $x,y$ in a bi-geodesic
$$
\phi^{\rm f}_{p}(x,y)~\le~  {4\over 3} \left[{r(1-p)\over p}\right]^{R_\GI  d_\GI(x,y)}
$$
where  $r= \left[2e\D^{2}\right]^{1/R_\GI}$.
\end{theo}

\\{\bf Remark}. The exponential decay of the finite connectivity functions in the supercritical phase
has been proved more than two decades ago, for the unit cubic lattice $ \mathbb{Z}^d$,
by  Chayes,  Chayes,  Newman \cite{CCN}  (see also \cite{CCGHS}). More recently Chen, Peres and Pete \cite{CPP} have shown that also non-amenable graphs (i.e. graphs with $C_\GI>0$) have
finite connectivity functions decaying exponentially.
We are not aware of any further generalization of such results in the literature.

\begin{theo}\label{supf}
Let $\GI=(\VU,\EE)$ a bounded degree graph which satisfies the hypothesis of Theorem \ref{criton2}, i.e. $\GI$ is such that $R_\GI>0$ and $P_\GI>0$. Then, as soon as
$ p \ge {4 \bar r\over 1+ \bar r} $ and any $x,y\in \VU$

$$
{1\over 3} \left[{(1-p)p^{R_\GI^{-1}}} \right]^{f_\GI(x,y)}~\le~\phi^f_{p}(x,y)~\le~{4\over 3} \left[{\bar r(1-p)\over p}\right]^{f_\GI(x,y)}
$$
\\where $\bar r= e^{1/P_\GI}\left[2\D^{2}\right]^{1/R_\GI}$.
\end{theo}

\\{\bf Remark}. Theorem \ref{supf} above shows  that in  a  graph $\GI$ with  $R_\GI>0$ and $P_\GI>0$, the contour distance $f_\GI(x,y)$ defined in \equ(fGI)
controls the decay of the two-point finite connectivity function in the supercritical phase in the sense that  $\phi^N_{p}(x,y)\propto (1-p)^{f_\GI(x,y)}$ as
$d_G(x,y)\to \infty$.  Therefore if $\GI$ is such that
$
\lim_{d_\GI(x,y)\to \infty }{f_\GI(x,y)/ d_\GI(x,y)}=0
$
the finite connectivity decays sub-exponentially. As an example, let us  consider the graph $\GI=(\VU,\EE)$ whose  vertex set $\VU$ is the subset  of $\Z^2$ given by $\VU=\{(n_1,n_2)\in \Z^2: n_1\ge0~{\rm and}~0\le n_2\le \ln(1+n_1)\}$
 and  whose edge set is formed by the nearest neighbors in $\VU$. It is easy to see that $R_\GI>0$ and $P_\GI>0$ and that $f_\GI((0,0),(n,0))= \lfloor\ln(1+n)\rfloor$ and so, by Theorem \ref{supf}, the connectivity function $\phi^N_{p}((0,0),(n,0))$ on $\GI$ is bounded below by  ${1\over 3}(1+n)^{-\a_1(p)}$ and bounded above by ${4\over 3}(1+n)^{-\a_2(p)}$ with
$\a_1(p)=|\ln[(1-p) p^{R^{-1}_\GI}]|$
and $\a_2(p)=|\ln[ \bar r(1-p)p^{-1}]|$.

\vskip.5cm
\numsec=4\numfor=1
\section{Proofs}\label{four}

We give here  the proofs of the four theorems stated in the previous section. We preliminarily introduce some few  notations.


We recall that, if $\g$ is a contour in $\GI$ and $x\in I_\g$, then
for any ray $\r=(V_\r, E_\r)$ in $\GI$ starting at $x$ we have that
$E_\r\cap \g\neq\emptyset$ (see e.g. Proposition 5.2 in \cite{PS8}). So,
for a fixed $x\in \VU$,  a fixed $\g\in \FF_\GI(x)$, and a fixed  geodesic ray  $\r$ starting at $x$, we  define
$e_{x}(\r,\g)$ as the first edge of $E_\r$, in the natural
order of the ray $\r$, which belongs to $\g$ and, for $n\in \mathbb{N}$, $x\in \VU$ and $\r$ geodesic ray in $\GI$,  let $r_n(x,\r)$ be the set
of edges of $\r$ defined as follows.
$$
r_n(x,\r)=\{e\in E_\r: \exists \g\in \FF^n_\GI(x) ~\mbox{such that }~
e=e_{x}(\r,\g)\}.\Eq(rn2)
$$

\subsection{\bf Proof of Theorems \ref{criton} and \ref{criton2}}
Suppose that $\GI$ has a bi-infinite geodesic $\d$. Choose $x$ to be a vertex of $\d$ and let $\r$ and $\r'$ be the two geodesic rays starting in $x$ such that
$\d=\r\cup\r'$.
Then, since the ray $\r$ is geodesic, we have $|r_n(x,\r)|\le \sup_{\g\in \FF^n_\GI}d_\GI(x, e_{x}(\r,\g))$ and since $\r\cup\r'$ is (bi-infinite) geodesic with $\r$ and $\r'$ starting at $x$,
$d_\GI(x, e_{x}(\r,\g))\le d_\GI(e_{x}(\r',\g), e_{x}(\r,\g))$ for any $\g\in \FF_\GI(x)$. Moreover, if $e,e'$ are any two vertices of $\g\in \FF_\GI$,
we have that $d_\GI(e, e')\le d_\GI^{\rm t}(\g)$, and, by \equ(iso1), $d_\GI^{\rm t}(\g)\le {|\g|/ R_\GI}$. So in the end we have
$$
|r_n(x,\r)|\le \sup_{\g\in \FF_\GI^n(x)}d_\GI(x, e_{x}(\r,\g))\le \sup_{\g\in \FF_\GI^n(x)} d_\GI(e_{x}(\r',\g), e_{x}(\r,\g))
$$
$$
\le \sup_{\g\in \FF_\GI^n(x)} d_\GI^{\rm t}(\g)\le {n\over R_\GI }\le e^{ n/ R_\GI}. \Eq(prim)
$$


\\Now observe that

$$
|\FF_\GI^n(x)|\le \sum_{e\in r_n(x,\r)}|\FF_\GI^n(e)| \le |r_n(x,\r)| \;\sup_{e\in E} \,|\FF_\GI^n(e)|.\Eq(pore)
$$
We  estimate $|\FF_\GI^n(e)|$, i.e. the number of contours of fixed cardinality $n$ containing a fixed edge $e$. To do this
we define a map $\t$ which associates to each contour $\g$ of cardinality $n$ and such that $e\in \g$ a tree $\t(\g)\subset \GI$ with edge set $E_{\t(\g)}$ such that $|E_{\t(\g)}|= d^{\rm t}_\GI(\g)$. Now, by hypothesis  contour  $R_\GI>0$. This implies that, for any contour $\g$ with cardinality $n$, $d^{\rm t}_\GI(\g)\le R_\GI^{-1}n$ and so we also have $|E_{\t(\g)}|\le R_\GI^{-1}n$. Moreover,  by definition $\g\subset E_{\t(\g)}$, and there are at most ${R_\GI^{-1}n\choose n}\le 2^{R_\GI^{-1}n}$ ways
to choose the set $\g$ in $E_{\t(\g)}$. So we get


$$
|\FF_\GI^n(e)|\le \sum_{\t\;{\rm tree \;in}\;G\atop |E_\t|=R_\GI^{-1}n,\, e\in \t}2^{R_\GI^{-1}n}
$$
and hence, using Euler's Theorem and the fact that $\GI$ has maximum degree $\D$,
$$
|\FF_\GI^n(e)| \le \left[2\D^{2}\right]^{n/R_\GI}\Eq(secon),
$$
uniformly in $e\in \EE$. In conclusion, by \equ(prim)-\equ(secon) we get
$$
|\FF_\GI^n(x)| \le~ r^n \Eq(paier)
$$
where
$$
r= \left[2e\D^{2}\right]^{1/R_\GI}\Eq(erre)
$$
and thus, by Proposition \ref{peia},  Theorem \ref{criton} is proved.

Choose now a vertex $x\in \VU$. Since $\GI$ is connected infinite and bounded degree, there exists a geodesic ray $\r$ starting at $x$. Let  $r_n(x,\r)$ the subset  of edges of $\r$
defined in  \equ(rn2). Since  $\r$ is geodesic and since, by hypothesis,  $P_\GI>0$, we have, for any  $\g\in \FF_\GI$, that
$|\g|\ge P_\GI\log[{\rm diam~}(I_\g)]$. So

$$
|r_n(x,\r)|\le \sup_{\g\in \FF_\GI^n(x)}{\rm diam~}(I_\g)
\le e^{n/P_\GI}.
$$
Since also $R_\GI>0$ we can use the bound  \equ(secon) together with bound \equ(pore) previously obtained to conclude that
$$
|\FF_\GI^n(x)| \le~ \bar r^n \Eq(fgnb)
$$
where now
$$
\bar r= e^{1/P_\GI} \left[2\D^{2}\right]^{1/R_\GI} \Eq(erbar)
$$
Theorem \ref{criton2} now follows once again from Proposition \ref{peia}.

\subsection{Proof of Theorem \ref{geoteo}}\label{sgeo}
Throughout this and the next subsections  we will denote shortly
$$
\l\equiv\l(p)=\frac{1-p}{p}
$$
We will also make use of the following definition: given an infinite  graph $\GI=(\VU, \EE)$, a sequence $\{V_N\}_{N\in\mathbb{N}}$ of
finite subsets of $\VU$ is said to {\it tend monotonically to $\VU$}, and we write $V_N\nearrow \VU$, if,
for all $N\in \mathbb{N}$, $V_N$ is connected, $V_{N}\subset V_{N+1}$, and $\cup_{N\in \mathbb{N}}V_N=\VU$.
We will denote shortly  $G_N=\GI[V_N]$ and  $E_N=\EE[V_N]$.

Let  $x,y$ be two vertices belonging to a bi-infinite geodesic of $\GI$. We choose a sequence $\{V_N\}_{N\in\mathbb{N}}$ tending monotonically to $\VU$ and suppose
$N$ so large that $\{x,y\}\in V_N\backslash \partial_v^{\rm int} V_N$.  Then, using the explicit representation \equ(restr) of the product measure $P_p$ restricted to $\O_N$,
we may define
the finite-volume finite connectivity functions
$$
\phi^{\rm f}_{p,N}(x,y)= \sum_{\om\in \O_{{N}}:~\exists g\,\,{\rm open\; cluster}\atop
\{x,y\}\subset V_g,~\partial g\subset E_N}p^{|O(\omega)|}(1-p)^{|C(\omega)|} \Eq(aio)
$$
So that
$$
\phi^{\rm f}_{p}(x,y)=\lim_{N\to \infty}\phi^{\rm f}_{p,N}(x,y)\Eq(limi)
$$
hence, by continuity of the product measure $P_p$, once we obtain an upper bound for $\phi^{\rm f}_{p,N}(x,y)$ uniformly in $N$, the same bound
also holds for the ``infinite-volume" limit $\phi^{\rm f}_{p}(x,y)$.

It is now easy to see that l.h.s. of \equ(aio) can be rewritten as
$$
\phi^{{\rm f},N}_{p}(x,y)=
{1\over Z_{N}(p)}\sum_{\om\in \O_{{N}}:~\exists g\,\,{\rm open\; cluster}\atop
\{x,y\}\subset V_g,~\partial g\subset E_N}
\l^{|C(\om)|}
\Eq(connef5)
$$

where

$$
Z_{N}(p)=\sum_{\omega\in \Omega_{N}}\l^{|C(\omega)|}
=p^{-|E_N|}.\Eq(bXi)
$$
So,
a configuration ${\omega\in \Omega_{N}}$ is given once we specify
the set of closed edges $C(\om)$ in $E_N$. If $C$ is a set of closed edges in $E_N$ we write
$C\odot \{x,y\}$ if there is a contour $\g\subset C$ such that $\g\odot \{x,y\}$. \\Then we can write

$$
Z_{N}(p)=\sum_{C\subset E_N}\l^{|C|}
=p^{-|E_N|}\Eq(bXi2)
$$
and
$$
\phi^{{\rm f},N}_{p}(x,y)=
{1\over Z_{N}(p)}
\sum_{{C\subset E_N\atop C\odot \{x,y\}}}
~\l^{|C|}.
\Eq(phifN)
$$
Hence
$$
\phi^{{\rm f},N}_{p}(x,y)\le
{1\over Z_{N}(p)}\sum_{\g\in \FF_\GI(x,y)}\l^{|\g|}
\sum_{{C\subset E_N\setminus \g\atop C\cup \g\odot \{x,y\}}}
~\l^{|C|}\le \sum_{\g\in \FF_\GI(x,y)}\l^{|\g|}.
$$
We now use the fact that $x,y$ belong to the bi-geodesic of $\GI$. Such a bi-infinite geodesic can be viewed as the union of two geodesic rays $\r$ and $\r'$ both starting at $x$
and such that $y\not\in V_\r$ and $y\in V_{\r'}$.
Fix now a contour $\g\odot \{x,y\}$.
Let,
$e(\r,\g)$ be the first edge of the ray $\r$ which belongs to $\g$ and let $e(\r',\g)$ be the first edge of the ray $\r'$ which belongs to $\g$.

Using the hypothesis that $R_\GI>0$,  by
definition \equ(iso1) we have that
$$
|\g| \ge R_\GI \cdot d^{\rm t}_{\GI}(\g) \ge R_\GI \cdot d_\GI(e(\r,\g),e(\r',\g))\ge R_\GI\cdot d_\GI(x,y)
$$
and so, using also the bound \equ(paier), we obtain

$$
\phi^{{\rm f},N}_{p}(x,y)\le \sum_{n\ge R_\GI d_\GI(x,y)} \sum_{\g\in \FF_\GI^n(x,y)}\l^{|\g|}\le \sum_{n\ge R_\GI d_\GI(x,y)} \l^n |\FF_\GI^n(x,y)|
$$
$$
\le \sum_{n\ge R_\GI d_\GI(x,y)} \l^n\sup_{x\in V} |\FF_\GI^n(x)| \le
\sum_{n\ge R_\GI d_\GI(x,y)} (r\l)^n = {{(r\l)^{R_\GI d_\GI(x,y)}}\over 1- r\l}.\Eq(lines)
$$
So we get, uniformly in $N$
$$
\phi^{{\rm f},N}_{p}(x,y)\le {4\over 3} (r\l)^{R_\GI d_\GI(x,y)},~~~~~~~~~~~~~~~~~~~{\rm for}~~\l\le {1\over 4r}
\hfill $$
which completes the proof of Theorem \ref{geoteo}.

\subsection{Proof of Theorem \ref{supf}}\label{ssu}

To obtain the upper and lower bounds for the finite connectivity, we work again at finite volume. We first obtain the upper bound which is easier. Proceeding analogously as we did in the proof of Theorem \ref{geoteo}, we rewrite $\phi^{{\rm f},N}_{p}(x,y)$
as the ratio \equ(phifN). Then
as above
$$
\phi^{{\rm f},N}_{p}(x,y)\le
{1\over Z_{N}(p)}\sum_{\g\in \FF_{G_N}(x,y)}\l^{|\g|}
\sum_{{C\subset E_N\setminus \g\atop C\cup \g\odot \{x,y\}}}
~\l^{|C|}\le \sum_{\g\in \FF_\GI(x,y)}\l^{|\g|}.
$$
Now recalling the Definition \ref{non root}, and following the same lines \equ(lines) with $\bar r$  defined in \equ(erbar) in place of $r$, we get for $\l\le {1\over 4\bar r}$ and uniformly in $N$,
$$
\phi^{{\rm f},N}_{p}(x,y)\le \sum_{n\ge f_\GI(x,y)} \sum_{\g\in \FF_\GI^n(x,y)}\l^{|\g|}\le 
{4\over 3} (\bar r\l)^{f_\GI(x,y)}.
$$

We now prove the lower bound of $\phi^{{\rm f},N}_{p}(x,y)$.
Let $\g_0$ be a minimum contour such that $\g_0\odot \{x,y\}$, that is, $|\g_0|=f_\GI(x,y)$, and $\g_0\subset E_N$ (we can always suppose $N$
sufficiently large to include that contour). We recall that by definition of contour, the set $E_N\setminus\g_0$ is partitioned in two disjoint sets $E_{\g_0}$ (the edge interior of $\g_0$)
and $E_N\setminus(\g_0\cup E_{\g_0})$ (the edge exterior of $\g_{xy}$) with $\GI[{I_{\g_0}}]=(I_{\g_0},E_{\g_0})$ being a connected graph. Let
$\t_0\subset E_{\g_0} $ be a minimal tree in $E_{\g_0}$ connecting the contour $\g_0$. By assumption (recall \equ(iso1)) we have that $|\t_0|\le R^{-1}_\GI|\g_0|$

Now, among all configurations $C$ of closed edges such that
$C\odot \{x,y\}$ there are those for which $C\supset \g_0$ and no subset of $C$ can separate $X$, and $C\cap \t_0=\0$ (i,e, all edges of $\t_0$ are open).
Then, summing only over these configurations, we get the lower bound
$$
\phi^{{\rm f},N}_{p}(x,y)\ge
{\l^{|\g_0|}\over Z_{N}(p)}\sum_{C\subset E_N\setminus \g_0\atop C\cap \t_0=\0,~C\cup \g_0 \odot \{x,y\}}\l^{|C|}.
$$
Now, since $E_N\setminus\g_0$ is the disjoint union of $E_N\setminus(\g_0\cup E_{\g_0})$ and $E_{\g_0}$, and observing that there is no restriction over the sum of closed edges in $E_N\setminus(\g_0\cup E_{\g_0})$ we have that
$$
\sum_{C \subset E_N\setminus \g_0\atop C\cap \t_0~=\0,~C\cup \g_0 \odot \{x,y\}}\l^{|C|}= Z_{E_N\setminus(\g_0\cup E_{\g_0})}
\sum_{C \subset E_{\g_0}\atop C\cap \t_0=\0,~C\cup \g_0 \odot \{x,y\}}\l^{|C|} ~= $$
$$
= p^{|\g_0|}{Z_N\over Z_{E_{\g_0}}} \sum_{C\subset E_{\g_0}\atop C\cap \t_0=\0,~C\cup \g_0 \odot \{x,y\}}\l^{|C|}~
= ~p^{|\g_0|}{Z_N\over Z_{E_{\g_0}}}\sum_{C\subset E_{\g_0}\setminus \t_0\atop C\cup \g_0 \odot \{x,y\}}\l^{|C|} ~=
$$
$$
= p^{|\g_0|}{Z_N\over Z_{E_{\g_0}}}\Bigg[Z_{ E_{\g_0}\setminus \t_0} -
\sum_{C\subset E_{\g_0}\setminus \t_0\atop C \otimes \{x,y\}}\l^{|C|}\Bigg]
$$
where $C \otimes \{x,y\}$ means that $C$ contains some contour $\g$ such that $\g\otimes \{x,y\}$.

Hence we get
$$
\phi^{{\rm f},N}_{p}(x,y)\ge
{(\l p)^{|\g_0|}\over Z_{E_{\g_0}}} \Bigg[Z_{ E_{\g_0}\setminus \t_0} -
\sum_{C\subset E_{\g_0}\setminus \t_0\atop C \otimes \{x,y\}}\l^{|C|}\Bigg]=
$$
$$
=
{(\l p)^{|\g_0|} p^{| \t_0|}} \Bigg[1 - {1\over Z_{ E_{\g_0}\setminus \t_0}}
\sum_{C\subset E_{\g_0}\setminus \t_0\atop C\otimes \{x,y\}}\l^{|C|}\Bigg]
\ge
$$
$$
\ge
\left[\l p^{1+R^{-1}_\GI}\right]^{|\g_0|} (1 - K_\l)
$$
where
$$
K_\l={1\over Z_{ E_{\g_0}\setminus \t_0}} \sum_{C\subset E_{\g_0}\setminus \t_0\atop C\otimes \{x,y\}}\l^{|C|}
$$
Now it is easy to get an upper bound for $K_\l$.
Indeed, since $\t_0$ is open in $E_{\g_0}$ and connects the boundary $\g_0$,
a configuration $C$ of closed edges can separate $\{x,y\}$ only if there is at least a contour surrounding either $x$ or $y$.
Hence
$$
K_\l ~\le~
{1\over Z_{ E_{\g_0}\setminus \t_0}} \sum_{\g\in \FF_{\GI}: \g\otimes \{x,y\}\atop \g\subset E_{\g_0}\setminus \t_0}\l^{|\g|}
\sum_{C\subset E_{\g_0}\setminus \t_0\atop C\cap \g=\0}\l^{|C|}~\le ~\sum_{\g\in \FF_{\GI}\atop \g\otimes \{x,y\}}\l^{|\g|} ~\le
$$
$$
\le~
2\sup_{x\in V}\sum_{\g\in \FF_{\GI}(x)}\l^{|\g|}~\le ~2
\sum_{n\ge 1}\l^n \sup_{x\in V}|\FF_\GI^n(x)|~~~~~~~~~~~
$$
Now, since $R_\GI>0$ and $P_\GI>0$ we can use the bound \equ(fgnb) to get
$$
K_\l~\le ~ 2
\sum_{n\ge 1}(\bar r\l)^n ~=~2 {\bar r\l\over 1-\bar r\l}
$$
and so $K_\l\le {2\over 3}$ as soon as $\l<{1\over 4\bar r}$. In conclusion, we have obtained, again uniformly in $N$,
$$
\phi^{{\rm f},N}_{p}(\{x,y\})\ge
{1\over 3} \left[\l p^{1+R^{-1}_\GI}\right]^{f_\GI(x,y)}
$$
as soon as  $\l<{1\over 4\bar r}$.

\section*{Acknowledgments}

We would like to thanks two anonymous referees who helped us, with their remarks and criticisms, to improve the presentation of the paper. We are also grateful to  Yuval Peres
for pointing out formula (A.3) in  reference \cite{CPP}. AP and RS have been partially supported by
CNPq and FAPEMIG (Programa de Pesquisador Mineiro).

\end{document}